\documentclass[journal]{IEEEtran}
\usepackage{graphicx}
\usepackage{epstopdf}
\usepackage{fancyhdr}
\usepackage{amsmath,amssymb,amstext}
\usepackage{subeqnarray}
\usepackage{cite}
\usepackage{hhline}
\usepackage{bm}
\usepackage{soul}
\usepackage{times,amsmath,amsfonts}
\usepackage{subfigure}
\usepackage{amsmath,amssymb}
\usepackage{graphicx,picture}
\usepackage{cite}
\usepackage{subfigure}
\usepackage{subeqnarray}
\usepackage{cases}
\usepackage{color}
\usepackage{overpic}
\usepackage{multirow}
\usepackage{booktabs}
\ifCLASSINFOpdf
\else
   \usepackage{graphicx}
   \graphicspath{{../eps/}}
   \DeclareGraphicsExtensions{.eps}
\fi
\usepackage{fancyhdr}
\usepackage{amsmath,amssymb,amstext}
\usepackage{subeqnarray}
\usepackage{cite}
\usepackage{hhline}
\usepackage{bm}
\usepackage{soul}
\usepackage{times,amsmath,amsfonts}
\usepackage{subfigure}
\usepackage{amsmath,amssymb}
\usepackage{graphicx,picture}
\usepackage{cite}
\usepackage{subfigure}
\usepackage{subeqnarray}
\usepackage{cases}
\usepackage{color}
\usepackage{overpic}  
\usepackage{multirow}
\usepackage{booktabs}
\usepackage{epstopdf}
\usepackage[ruled, linesnumbered]{algorithm2e}

\definecolor{r}{rgb}{1,0,0}
\definecolor{b}{rgb}{0,0,1}
\definecolor{k}{rgb}{0,1,1}
\usepackage{upgreek}

\newcounter{saveeqn}%

\allowdisplaybreaks
\DeclareMathSymbol{\Phi}{\mathord}{letters}{8}

\begin{document}

\title{Channel Knowledge Map Enabled Low-Complexity Dynamic Radio Environment Reconstruction}

\author{
 \IEEEauthorblockN{Yujun Lin, Zhiqiang Xiao, Hao Wu, Xiaoqiang Qiao, Fayu Wan, and Tao Zhang
 }

\thanks{
Y. Lin is with the Nanjing University of Information Science and Technology, Nanjing, 210044, China, and also with the Sixty-Third Research Institute, National University of Defense Technology, Nanjing, 210007, China (email: 17268233472@163.com).

Z. Xiao, H. Wu, X. Qiao and Tao Zhang are with the Sixty-Third Research Institute, National University of Defense Technology, Nanjing, 210007, China (e-mail: zhiqiang\_xiao@nudt.edu.cn; whao1983@126.com; qxq0527@163.com; ztcool@126.com).

F. Wan is with the Nanjing University of Information Science and Technology, Nanjing, 210044, China (email: fayu.wan@nuist.edu.cn).
}
}

\maketitle

\begin{abstract}
Accurate and timely radio environment reconstruction is important but challenging under particularly dynamic transmitter configurations.
The conventional methods such as compressed sensing (CS), Kriging method or U-Net typically require environment measurements and reconstruction overhead for radio environment updating as the transmitter locations or radiation patterns change.
In this paper, we propose a novel channel knowledge map (CKM)-enabled dynamic radio environment reconstruction method for efficient radio map updating.
Specifically, the recently proposed CKM can store reusable path-level propagation knowledge that is decoupled from the transmitter-side radiation characteristics.
We can leverage CKM for lightweight forward radio map generation as the transmitter locations and radiation patterns are known, without requiring new target-map measurements.
Simulation results show that the proposed method outperforms CS, Kriging, and U-Net in reconstruction accuracy and exhibits strong robustness performance under dynamic transmitter configurations, which demonstrates the potential of the proposed method for flexible and efficient radio environment reconstruction in dynamic wireless networks.
\end{abstract}

\begin{IEEEkeywords}
Channel knowledge map, dynamic transmitter configuration,
radio environment reconstruction, radio map, received power.
\end{IEEEkeywords}

\section{Introduction}

\IEEEPARstart{R}{adio} environment reconstruction aims to characterize the spatial distribution of radio parameters over a target region, and serves as an important basis for spectrum management, wireless network planning, localization, and environment-aware optimization \cite{Romero2022}.
A typical representation of the radio environment is the radio map (RM), which describes location-dependent quantities such as power spectral density, path loss, and received power.
In this paper, we focus on received signal power-based RM construction.
As wireless systems evolve toward denser and more heterogeneous deployments, the regional radio environment is increasingly affected by changes in transmitter locations, transmit
powers, radiation patterns, and the number of transmitters.
Therefore, accurate and timely radio environment reconstruction becomes both more valuable and more challenging.

Existing RM construction methods can be broadly divided into source-aware prediction and source-agnostic inverse reconstruction.
Source-aware methods predict the RM from known transmitter information and environmental geometry, with ray tracing and learning-based RadioUNet as representative examples \cite{Yun2015RayTracing,Levie2021}.
Source-agnostic methods recover the RM from limited spatial measurements, including Kriging interpolation \cite{DallAnese2011Kriging}, compressed sensing (CS) \cite{Shen2022}, and CNN-based reconstruction \cite{Teganya2022}.
Despite their differences, these methods usually regard the RM under a given transmitter configuration as the direct construction target.

Meanwhile, the channel knowledge map (CKM) has emerged as a promising paradigm for organizing and exploiting location-specific channel knowledge \cite{Zeng2021}.
We focus on scenarios in which the propagation environment remains quasi-static over the period of interest, while transmitter configurations vary more frequently.
In this setting, reusable propagation knowledge can be combined with configuration-dependent transmitter characteristics to generate updated RMs.

Motivated by the above insight, this paper develops a CKM-based framework for dynamic radio environment reconstruction, represented by received power-based RM construction.
Different from conventional methods that directly construct configuration-specific received power maps, the proposed approach stores reusable propagation knowledge in a CKM and synthesizes the desired received power map by combining the queried CKM parameters with transmitter-side radiation characteristics.
In this way, transmitter relocation, power adjustment, radiation-pattern change, and transmitter-number variation can be handled through CKM querying and lightweight forward computation, without new field measurements, repeated map reconstruction, or per-configuration training.

The remainder of this paper is organized as follows.
Section~II introduces the system model, conventional RM construction methods, and the CKM-enabled reconstruction framework.
Section~III presents the proposed CKM construction and CKM-based RM generation method.
Section~IV reports the simulation results, and Section~V concludes the paper.

\section{System Model}\label{sec:system_and_method}

\begin{figure}[!t]
    \centering
    {
    \includegraphics[width=0.4\textwidth]{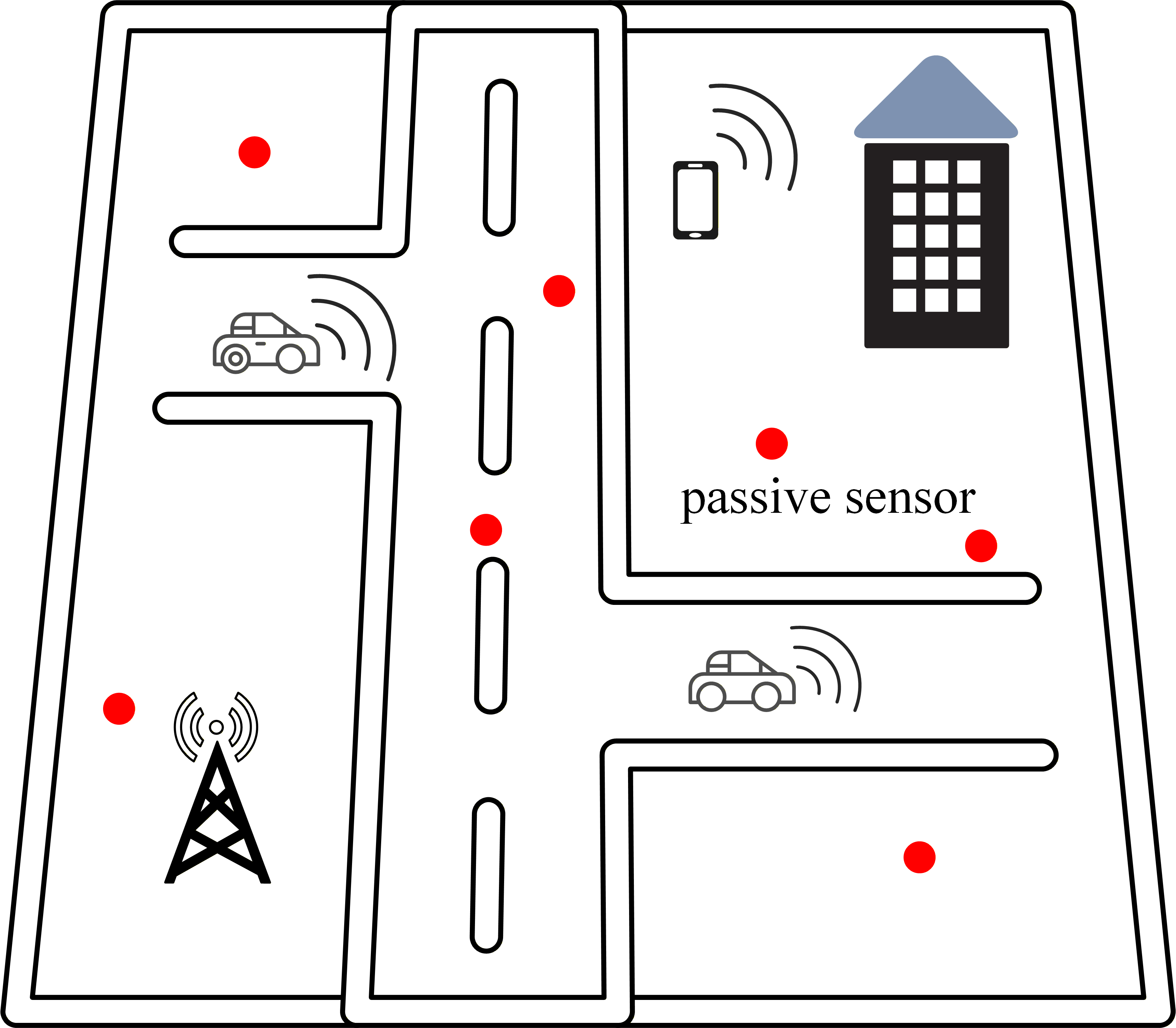}
  \caption{An illustration of a dynamic radio environment with heterogeneous transmitters and passive sensors.}\label{system model}
    }\vspace{-15pt}
\end{figure}
In this paper, we consider a radio environment reconstruction problem, which refers to reconstructing the received power distribution over a geographic area of interest.
As illustrated in Fig.~\ref{system model}, such areas contain multiple heterogeneous wireless signal transmitters, including fixed wireless hotspots, mobile users, and vehicles.
The transmitted signals propagate through complex wireless channels and intricately intertwine within the area of interest.
As a result, the radio environment is complex and dynamic, posing significant challenges for accurate and timely reconstruction.

For ease of exposition, we consider a quasi-static propagation environment over the time duration of interest and assume that the channel variations are governed solely by the TX-RX location geometry.
Accordingly, by collecting the received power over the spectrum of interest $f\in\mathcal{F}$, across all $Q_r$ sensors, the received power distribution over the area can be obtained as
\begin{equation}\label{eq:psd_map}
\mathcal{M} = \{Z(\mathbf{r},f)| \mathbf{r}\in\mathcal{Q}_r, f\in\mathcal{F}\},
\end{equation}
where $\mathbf{r}\in\mathbb{R}^{2}$ denotes the location of a passive sensor in horizontal plane, while $\mathcal{Q}_r$ and $\mathcal{F}$ represent the sets of sensor locations and spectrum of interest, respectively.

However, due to hardware cost and implementation constraints, directly measuring the received signal power over a wide area and across a broad spectrum is challenging.
The concept of a RM has recently been proposed, which aims at providing a continuous representation of the received signal power distribution over the geographic area of interest across the target spectrum.

\subsection{Conventional Radio Environment Reconstruction Methods}
The conventional RM construction methods can be loosely classified into two categories, i.e., source-aware prediction and source-agnostic inverse reconstruction.

For source-aware prediction methods \cite{Li2025,Zhang2020}, the transmitter (TX) location and its radiation parameters are prior known, and the objective is to forward predict the RM from the propagation environment.
To this end, deep neural networks (DNNs) are often employed to learn the implicit mapping from the propagation environment and TX configuration to the spatial spectral received power distribution.
Let $\mathcal{D}_{\bm\theta}(\cdot)$ be the parametric function to be learned, such that
\begin{equation}
\hat{Z}(\mathbf{r}_{i},f)=\mathcal{D}_{\bm{\theta}}(\mathbf{r}_{i},f,\{{G}_t,\mathbf{q}_t, t\in\mathcal{T}\},\mathbf{E}),
\end{equation}
where $\boldsymbol{\theta}$ denotes the learnable parameters,  $\hat{Z}(\mathbf{r}_{i}, f)$ is the predicted received signal power, ${G}_t$ and $\mathbf{q}_t$ represent the radiation parameters and location of the $t$th TX for $t\in\mathcal{T}$, respectively, with $\mathcal{T}\triangleq\{1,\cdots,T\}$ denoting the index set of transmitters, and $\mathbf{E}$ is the propagation environment.
The network is typically trained by minimizing the mean squared error (MSE) between the predicted and ground-truth received power over a set of training samples $\{Z_i\}_{i=1}^{N}$, i.e.,
$\mathcal{L}(\boldsymbol{\theta}) = \frac{1}{N} \sum_{i=1}^{N} \left|\hat{Z}(\mathbf{r}_{i}, f) - Z_i \right|^2
$.
As a result, the mapping from the environment geometry and TX configuration to the spatial spectral received power distribution can be obtained.

Alternatively, RM construction can be formulated as an inverse problem.
Specifically, to recover the received power distribution over all $M$ spatial-spectral grids from a limited set of $N$ measurements, with $N\ll M$, an inverse problem can be formulated as
\begin{equation}
\mathbf{y} = \bm{\Phi}\mathbf{x} + \mathbf{n},
\label{eq:bcs_model}
\end{equation}
where $\mathbf{x} \in \mathbb{C}^{M \times 1}$ denotes the discretized received power to be recovered, $\mathbf{y} \in \mathbb{C}^{N\times 1}$ represents the $N$ collected measurements, $\bm{\Phi} \in \mathbb{R}^{N \times M}$ is the sampling matrix, and $\mathbf{n}\in\mathbb{C}^{N\times1}$ is measurement noise.
Since $\bm{\Phi}$ is typically rank-deficient, the problem is underdetermined and difficult to solve directly \cite{li2020cs_overview}.
To address this challenge, extensive research efforts have been devoted to this field, including Kriging-based spatial interpolation, CS-based sparse recovery \cite{tropp2007omp}, and CNN-based radio-map reconstruction.

The aforementioned methods primarily focus on the RM construction under a given TX configuration.
However, even in a quasi-static propagation environment, any changes in TX locations or radiation characteristics can lead to significant variations of the RM.
As a result, the entire map reconstruction or regional updating is necessary, which inevitably incurs substantial operational overhead.

To address this limitation, in this paper, we propose a low-complexity dynamic spectrum environment reconstruction framework based on the recently proposed CKM, which decouples the propagation environment from TX configuration by leveraging CKM as an intermediate representation.

\subsection{Proposed CKM-Enabled Reconstruction Method}
Specifically, consider an any-to-any (X2X) CKM.
For each TX-RX location pair within the area of interest, the CKM can be regarded as a mapping $\mathcal{K}$ from the location pair $(\mathbf{r},\mathbf{q}_t)$ to the channel knowledge vector $\mathbf{c}\in\mathbb{C}^{J}$, i.e.,
\begin{equation}
\mathcal{K}:\ (\mathbf{r},\mathbf{q}_t)
\ \longmapsto\ \mathbf{c}(\mathbf{r},\mathbf{q}_t),
\end{equation}
where $\mathbf{c}(\mathbf{r},\mathbf{q}_t)$ captures propagation knowledge with the dimension of ${J}$.
In this work, we consider that the CKM stores a set of the path state information (PSI) \cite{PSI2025} for the given location pair $(\mathbf{r},\mathbf{q}_t)$, i.e.,
\begin{equation}
\mathbf{c}(\mathbf{r},\mathbf{q}_t)
=
\left\{
\lvert \alpha_{l,t} \rvert,\ \angle \alpha_{l,t},\ \tau_{l,t},\ \phi_{l,t}
\right\}_{l=1}^{L},
\label{eq:ckm_content}
\end{equation}
where $L$ is the number of significant multi-paths, $\lvert \alpha_{l,t} \rvert$ and $\angle \alpha_{l,t}$ denote the amplitude and phase of the $l$th propagation path, respectively, $\tau_{l,t}$ is the corresponding propagation delay, and $\phi_{l,t}$ is the angle-of-departure (AoD) of the $l$th path.
Here, we consider a millimeter wave (mmWave) frequency band radio environment reconstruction, where the number of significant propagation paths per location pair is typically small.
Here, we assume that all location pairs have the same number of multi-paths $L$.
This in fact loses no generality as we could simply add virtual paths with zero gain for any location pair with fewer multi-paths.

\begin{figure}[!t]
    \centering
    \includegraphics[width=0.45\textwidth]{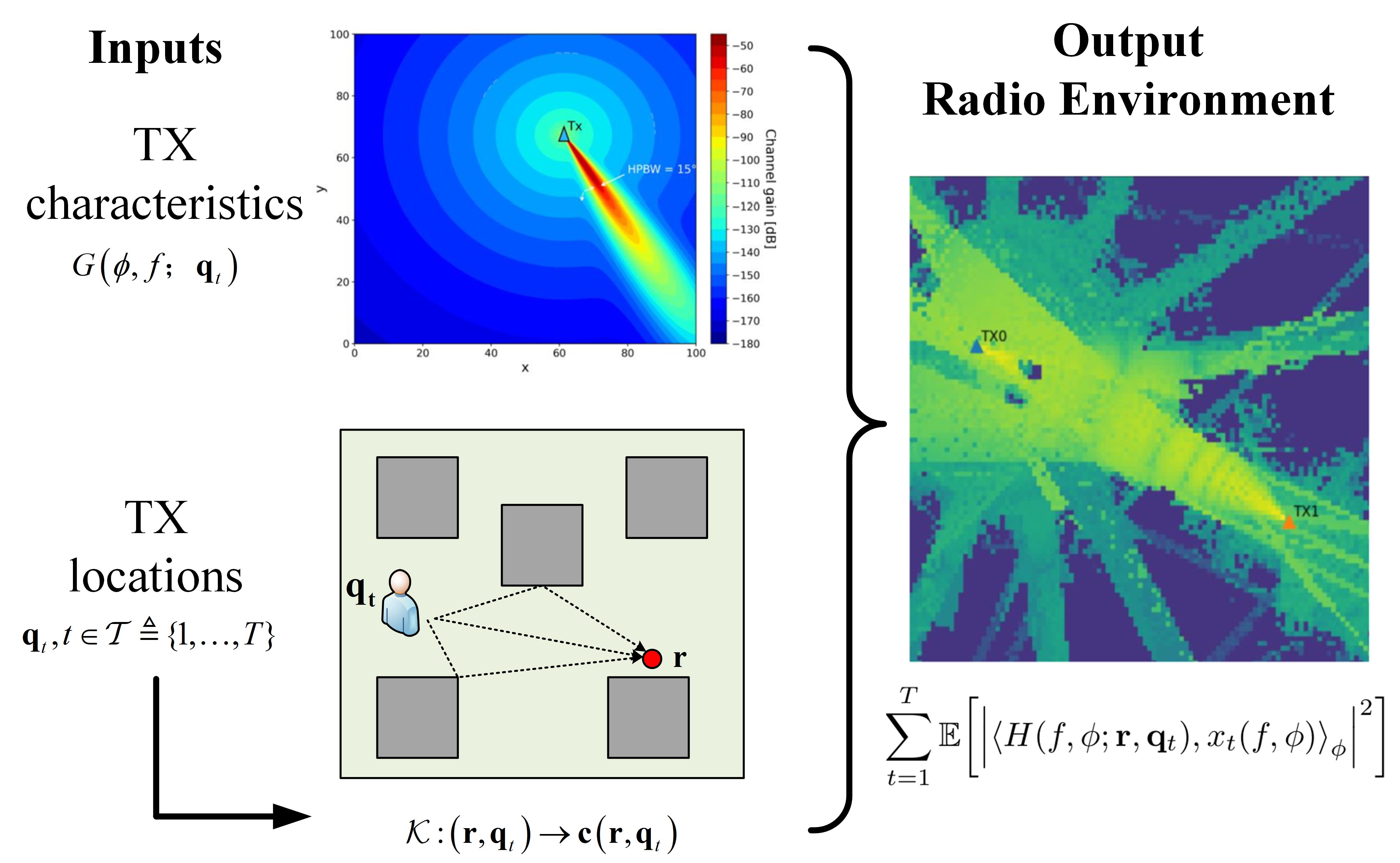}
    \caption{The proposed CKM-based radio environment construction method.}
    \label{fig:ckm}
\end{figure}

For any given location pair $(\mathbf{r},\mathbf{q}_t)$, with the obtained channel knowledge in \eqref{eq:ckm_content}, the spatial-frequency channel response can be expressed as
\begin{equation}
H(f,\phi;\mathbf{r},\mathbf{q}_t)
=
\sum_{l=1}^{L}
\alpha_{l,t}
e^{-j2\pi f\tau_{l,t}}
\delta(\phi-\phi_{l,t}),
\label{eq:henv}
\end{equation}
where $\alpha_{l,t}\triangleq \lvert \alpha_{l,t}\rvert e^{j\angle \alpha_{l,t}}$ denotes the complex gain of the $l$th propagation path.
Therefore, as shown in Fig.~\ref{fig:ckm}, as the TX configurations are prior known, the corresponding noise-free received power $Z(\mathbf{r},f)$ can be calculated immediately as

\begin{equation}\label{eq:rm_definition}
\begin{aligned}
Z(\mathbf{r},f)
&=
\mathbb{E}\!\left[
\left|
\sum_{t=1}^{T}
\left\langle
H(f,\phi;\mathbf{r},\mathbf{q}_t),
x_t(f,\phi)
\right\rangle_{\phi}
\right|^2
\right]
\\
&\overset{(a)}{=}
\sum_{t=1}^{T}
\mathbb{E}\!\left[
\left|
\left\langle
H(f,\phi;\mathbf{r},\mathbf{q}_t),
x_t(f,\phi)
\right\rangle_{\phi}
\right|^2
\right],
\end{aligned}
\end{equation}
where $(a)$ follows from the assumption of incoherent signal superposition and $x_t(f,\phi)$ denotes the transmit signal of the $t$th TX, whose expression is to be derived later in Section III.
Note that the angular coupling integral is used in \eqref{eq:rm_definition}, which is defined as
$
\left\langle u,v\right\rangle_{\phi}
\triangleq
\int_{-\pi}^{\pi}
u(\phi)v(\phi)\,\mathrm{d}\phi .
$

\section{Proposed CKM-Enabled Radio Environment Reconstruction Method}

\subsection{CKM Construction}

In this paper, the CKM composed of the PSI for each TX-RX location pair can be constructed through data acquisition followed by spatial prediction.
Specifically, the region of interest can be first discretized into several spatial grids.
Then, the channel measurement can be conducted offline using a dedicated test device that traverses all grids to obtain the PSI as given in \eqref{eq:ckm_content}, or in an opportunistic manner by leveraging the channel estimation from other users.
Subsequently, when the PSI for the finite TX-RX location pairs have been obtained, many interpolation techniques such as K-nearest neighbors (KNN), inverse-distance weight (IDW), and the Kriging algorithms, or deep neural networks can be trained for CKM construction~\cite{Tutorial}.

Note that, different from the radio map, which may vary rapidly as the transmit power or radiation pattern change, the CKM depends only on the propagation environment that evolves on a much slower timescale.
Therefore, we can construct the CKM once and reuse it for dynamic radio environment reconstruction as the TX configurations change, eliminating the need for repeated site-specific measurements.

\subsection{Propagation-Aware Radio Environment Reconstruction}\label{sec:radiation}
\begin{figure*}[!t]
\centering
\includegraphics[width=\textwidth]
{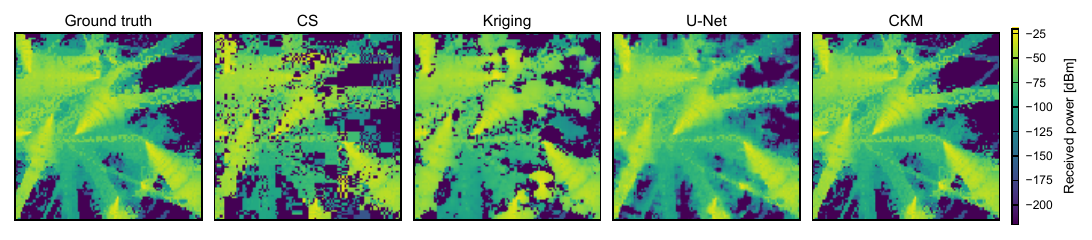}
\caption{Comparison of the radio environment reconstruction accuracy for different methods.}
\label{fig:power_map_comparison}
\end{figure*}

With the CKM, for a given TX-RX location pair $(\mathbf{r},\mathbf{q}_t)$, the PSI and the corresponding channel response can be obtained from \eqref{eq:ckm_content} and \eqref{eq:henv}, respectively.
Therefore, according to \eqref{eq:rm_definition}, as the spatial-frequency radiation pattern for all transmitters, i.e., $\{x_t(f,\phi)\}_{t=1}^{T}$, are known, the radio environment at any location $\mathbf{r}$ can be directly calculated.

Specifically, for ease of exposition, the radiation pattern for the $t$th TX can be modeled as
\begin{equation}\label{tx model}
x_t(f,\phi)
=
\sqrt{G(\phi-\phi_{t})}s_t(f),
\end{equation}
where $G(\phi-\phi_{t})$ denotes the angle-dependent radiation pattern and $s_t(f)$ represents the transmit information-bearing symbol at the frequency $f$ and $s_t(f)\sim\mathcal{CN}(0,P_t)$, with $P_t$ being the average transmit power.
Here, we assume that $G(0)$ is the maximum gain, with $\phi_t\in[-\pi,\pi)$ denoting the orientation angle of the $t$th TX.

By substituting \eqref{eq:henv} and \eqref{tx model} into \eqref{eq:rm_definition}, we obtain
\begin{equation}
\begin{aligned}
Z(\mathbf{r},f)
&=
\sum_{t=1}^{T}
\mathbb{E}\!\left[
\left|
\left\langle
H(f,\phi;\mathbf{r},\mathbf{q}_t),
x_t(f,\phi)
\right\rangle_{\phi}
\right|^2
\right]
\\
&=
\sum_{t=1}^{T}
P_t
\left|
\sum_{l=1}^{L}
\alpha_{l,t}
e^{-j2\pi f \tau_{l,t}}
\sqrt{G(\phi_{l,t}-\phi_t)}
\right|^2 .
\end{aligned}
\label{eq:rm_power}
\end{equation}
Finally, the complete radio environment can be reconstructed according to \eqref{eq:rm_power} over all  $\mathbf{r}\in\mathcal{Q}_r$.

Note that compared with conventional methods, the proposed CKM-enabled radio environment reconstruction framework enjoys the following advantages:

(i) \emph{Low-overhead radio environment updating.}
With the CKM, the radio environment can be efficiently updated to accommodate dynamic changes with the TX configurations, including the transmit power, radiation pattern, and the number of transmitters, without requiring additional measurements.

(ii) \emph{CKM as a reusable propagation foundation.}
Once constructed, the CKM serves as a fundamental propagation knowledge base for radio environment reconstruction, enabling efficient generation of diverse radio maps under arbitrary TX deployments within the same environment.

\section{Simulation Results}

\begin{table}[t]
\caption{Main simulation parameters.}
\label{tab:simulation_parameters}
\centering
\begin{tabular}{lc}
\hline
Parameter & Value \\
\hline
\multicolumn{2}{l}{\textit{Scene and map configuration}} \\
\multicolumn{2}{l}{\quad Reconstruction-accuracy experiment} \\
Area / grid spacing
& $500\times500~\mathrm{m}^2$ / $5$~m \\
Grid size & $101\times101$ \\
\multicolumn{2}{l}{\quad Robustness experiment} \\
Area / grid spacing
& $1000\times1000~\mathrm{m}^2$ / $10$~m \\
Grid size & $101\times101$ \\
Carrier frequency & $28$~GHz \\
TX/RX height & $1.6$~m \\
\hline
\multicolumn{2}{l}{\textit{Propagation and antenna configuration}} \\
Samples per source & $10^4$ \\
Maximum path depth & $4$ \\
TX pattern & Gaussian, $15^\circ$ HPBW \\
TX gain range & $-30$ to $27.79$~dBi \\
Transmit power & $27$, $30$, or $33$~dBm \\
Boresight direction & Uniform in $[-\pi,\pi)$ \\
\hline
\multicolumn{2}{l}{\textit{Baseline configuration}} \\
Observation ratio & $\rho=60\%$ \\
CS & DCT-OMP ($6\times6$, $K\leq5$) \\
Kriging & Ordinary ($32$ neighbors, $120$~m) \\
U-Net data split & $200/30/20$ \\
\hline
\end{tabular}
\end{table}

In this section, experiments are conducted using Sionna RT 1.2.1 in \texttt{sionna.rt.scene.etoile} and \texttt{sionna.rt.scene.munich} scenes \cite{AitAoudia2025Sionna}.
The evaluated areas in the Etoile and Munich scenes cover $500\times500~\mathrm{m}^2$ and $1000\times1000~\mathrm{m}^2$, respectively, and are sampled at grid spacings of $5$~m and $10$~m, yielding $101\times101$ grid points in both experiments.
The carrier frequency is $28$~GHz, and the TX and RX heights are both set to $1.6$~m.

Here, we employ the built-in \texttt{PathSolver} of Sionna RT to compute the propagation paths between the TXs and RXs, configured with $10^4$ rays sampled per source and a maximum propagation depth of four.
The RX is isotropic, while each TX employs an azimuth-dependent Gaussian power-gain pattern given by
\begin{equation}
G(\phi)
=
G(0)\exp\left(
-\frac{4\ln 2}{\phi_{\mathrm{3dB}}^2}\phi^2
\right),
\label{eq:gaussian_pattern}
\end{equation}
where $G(0)$ denotes the maximum antenna gain and $\phi_{\mathrm{3dB}}$ denotes the half-power beamwidth, which are set to $27.79$~dBi and $15^\circ$, respectively.

The ground-truth map is generated by direct ray tracing, whereas the proposed method reconstructs the map using the stored PSI and the known TX configuration.
The conventional CS, Kriging, and U-Net are selected as the benchmark, which use the same target-map observations, and the target-map observation ratio is set to $\rho=60\%$.
The U-Net training, validation, and internal-test sets contain 200, 30, and 20 maps, respectively, whose TX configurations are disjoint from the 100 evaluation cases.
The main parameter settings are summarized in Table~\ref{tab:simulation_parameters}.
\begin{table}[t]
\centering
\caption{Reconstruction accuracy for 10-TX received-power maps.}
\label{tab:reconstruction_accuracy}
\begin{tabular}{lcccc}
\hline
& MSE $\downarrow$
& RMSE $\downarrow$
& NMSE $\downarrow$
& SSIM $\uparrow$ \\
\hline
CS & 0.0417 & 0.2041 & 0.1121 & 0.1933 \\
Kriging & 0.0300 & 0.1733 & 0.0808 & 0.3329 \\
U-Net & 0.0022 & 0.0472 & 0.0060 & 0.7384 \\
CKM & \textbf{0.0003} & \textbf{0.0161} & \textbf{0.0007} & \textbf{0.9837} \\
\hline
\end{tabular}
\end{table}

\subsection{Reconstruction Accuracy}
The radio environment reconstruction accuracy is first evaluated in \texttt{sionna.rt.scene.etoile} scene using 100 independent cases, each comprising ten outdoor TXs with different locations, transmit powers, and boresight directions.

Following the normalized map-evaluation metrics in \cite{fu2026ckmdiff}, we evaluate the reconstructed power maps using mean squared error (MSE), root mean squared error (RMSE), normalized mean squared error (NMSE), and structural similarity index measure (SSIM).
The MSE and NMSE are computed over all valid outdoor points across the 100 cases, while RMSE is the square root of MSE, and SSIM ($\mathrm{data\_range}=1$) is computed per map and then averaged.

As shown in Table~\ref{tab:reconstruction_accuracy}, the proposed CKM-based method achieves the best performance for all four metrics.
Specifically, the RMSE of the proposed CKM-enabled method is less than half that of U-Net and substantially lower than those of CS and Kriging, while the SSIM of 0.9837 indicates it better preservation of the spatial power-distribution structure.

Fig.~\ref{fig:power_map_comparison} provides a qualitative comparison for the reconstruction accuracy for different methods.
It can be observed that CS and Kriging exhibit larger reconstruction errors and recover less spatial detail in this example, while U-Net captures the overall distribution but introduces visible smoothing.
In contrast, the CKM result closely follows the ground truth in both coverage structure and local power variation.

\subsection{Robustness and Overhead Analysis}
To further investigate the robustness of the proposed method, we evaluate its performance under varying TX counts.
Specifically, to mitigate the potential spatial smoothing caused by increasing TX counts, we replace the original scene with the larger \texttt{sionna.rt.scene.munich} scene and vary the number of active TXs as $|\mathcal{T}|\in\{2,\ldots,14\}$, while keeping all other simulation settings unchanged.

\begin{figure}[!t]
\centering
\includegraphics[width=\columnwidth]
{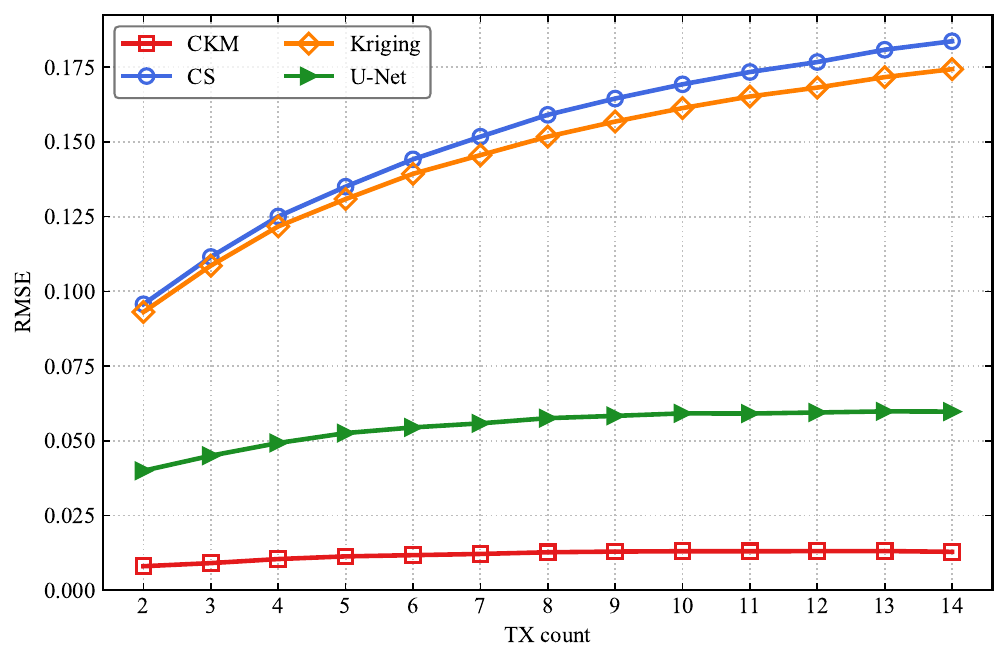}
\caption{RMSE versus the TX count.}
\label{fig:reconfiguration_observation_cost}
\end{figure}

As shown in Fig.~\ref{fig:reconfiguration_observation_cost}, CKM consistently achieves the lowest RMSE and exhibits only limited performance variation across all evaluated TX counts.
However, both CS and Kriging suffer from a pronounced rise in RMSE as the TX count increases.
Although U-Net performs relatively better than these two, it still fails to preserve its accuracy as the scenario complexity grows.
In contrast, the proposed CKM-enabled method demonstrates remarkable robustness.

Beyond its robustness, the proposed method also reduces the measurement and adaptation overhead associated with RM updating.
CS, Kriging, and U‑Net demand $60\%$ fresh coverage per TX configuration, with U-Net often needing retraining when TX configuration deviates significantly from training data.
In contrast, the proposed CKM-enabled method generates an accurate RM through CKM querying and lightweight forward computation, without acquiring new target-map measurements or retraining a prediction model.
This property is particularly valuable for complex and dynamic radio environment reconstruction.

\section{Conclusion}
\label{sec:conclusion}
This paper proposed a CKM-based framework for reconstructing received-power RMs in dynamic radio environments.
By decoupling reusable propagation knowledge from TX-specific radiation characteristics, the proposed method updates the RM through CKM querying and forward computation without requiring additional target-map measurements.
Simulation results demonstrate that the proposed method achieves superior reconstruction accuracy among the considered methods and exhibits strong robustness performance across varying numbers of active TXs.
More importantly, once constructed, the CKM serves as a reusable propagation foundation that accommodates changes in TX power, radiation pattern, and number, enabling low-overhead generation of diverse RMs within the same physical environment.
Future work will validate the proposed framework in real-world complex environments and investigate its integration with practical measurement-driven CKM construction.

\bibliographystyle{IEEEtran}
\bibliography{reference}

\end{document}